\documentclass[a4paper]{jpconf}
\usepackage{graphicx}
\usepackage{bm,color}
\begin{document}
\title{Co-existence of states in quantum systems}

\author{Yoritaka Iwata}

\address{School of Science, The University of Tokyo, Hongo 7-3-1, 113-0033 Tokyo}

\ead{iwata@cns.s.u-tokyo.ac.jp}

\begin{abstract}
Co-existence of different states is a profound concept, which possibly
underlies the phase transition and the symmetry breaking.
Because of a property inherent to quantum mechanics (cf. uncertainty), the
co-existence is expected to appear more naturally in quantum-microscopic systems
than in macroscopic systems.
In this paper a mathematical theory describing co-existence of states in quantum
systems is presented, and the co-existence is classified into 9 types.
\end{abstract}

\section{Introduction}
The boundary-value problem of nonlinear partial differential equation of elliptic-type:
\begin{equation} \label{mastereq} \left\{ \begin{array}{ll}
- \nabla^2 u - m u - V(u) u  = 0 \quad {\rm in}~ \Omega, \vspace{2.5mm}\\
 u = 0 \quad {\rm on}~ \partial \Omega,
\end{array} \right. \end{equation}
is studied, where $m$ is a real number, and $\Omega \in R^3$ is a closed
domain with a sufficiently smooth boundary.  
The unknown complex function $u$ consists of the unknown state $\psi$ and the reference state ${\bar \psi}$: 
\[ u = \psi - {\bar \psi}, \]
where ${\bar \psi}$ (corresponding to a generalized concept of the vacuum) is not necessarily a solution of Eq.~(\ref{mastereq}), although the most simplest case ${\bar \psi}=0$ (the simplest vacuum) satisfies Eq.~(\ref{mastereq}). 
Let a part of the inhomogeneous term $V(u)$, whose spectral set is assumed to be included in a
real axis, satisfy 
\begin{equation} \label{inhomcond} \begin{array}{ll} 
\partial_{u} (V(u) u)|_{u=0} = V_L. 
\end{array} \end{equation}
For the simplicity $V_L$, which corresponds to the signed strength of linearized
interaction being independent of $u$, is assumed to be a real number.
As is readily seen, the function $u = \psi - {\bar \psi} = 0$ is always a solution of this problem (refer to the
trivial solution).
In this sense let us imagine a simple case when ${\bar \psi} = 0$, and
then the emergence of a solution $\psi$ from another solution ${\bar \psi}=0$ is true if $u \ne 0$ is the solution of Eq.~(\ref{mastereq}).
Here we seek the non-trivial solution $u \ne 0$ ($\psi \ne {\bar \psi}$) to Eq.~(\ref{mastereq}).
The corresponding situation is nothing but the co-existence of different states $\psi$ and ${\bar \psi}$.

Equation ~(\ref{mastereq}) is associated with the stationary problem of nonlinear
Schr\"odinger equations as well as nonlinear Klein-Gordon equations.
In the context of Klein-Gordon equations, it is possible to associate $\sqrt{-m}$ with
the mass (if $m < 0$).
Note that the statistical property inherent to many-body system, which might
bring about rather interesting physical properties, is not taken into account
in order to see the most fundamental properties associated with the
co-existence in both nonlinear
Schr\"odinger equations and nonlinear Klein-Gordon equations.

\section{Theory describing the co-existence}
\subsection{Mathematical settings}
Let $X$ and $Y$ be functional spaces
\[ \begin{array}{ll}
X = W_0^{1,2}(\Omega) \cap W^{2,2}(\Omega), \vspace{1.5mm} \\
Y = L^2(\Omega)
\end{array} \]
respectively (for mathematical notation, see \cite{80yosida}).
An inclusion relation $X \subset Y$ is true.
For $u \in X$,  a mapping $f:R^1 \times X  \to Y$ is defined by
\[
f(\lambda, u) := - \nabla^2 u - m u - V(u) u.
\]
The original master equation is written by $f(\lambda, u)= 0$.
Since the trivial solution $u = 0$ always exists, $f(\lambda, 0)= 0$ is satisfied. 
According to the Sobolev embedding theorem $- \nabla^2$
is a $C^2$-mapping from $X$ to $Y$, where the detail setting of $V(u)$ is necessary to know the regularity of the mapping $f$. 
The space $W_0^{1,2}(\Omega)$ denotes all the functions included in $W^{1,2}(\Omega)$ satisfying $u|_{\partial \Omega} = 0$.

\subsection{Linearized analysis} \label{linear}
Linearized problem is derived.
The Fr${\acute {\rm e}}$chet derivative of $f(\lambda, u)$ is calculated as
\begin{equation}  \label{linemastereq} 
f_{u}(\lambda, 0) [u] = - \nabla^2 u - \lambda u = 0,
\end{equation}
where $\lambda = m + V_L$.
This corresponds to the master equation for the linearized eigen-value problem.
It is well known that the linearized problem (with the Dirichlet boundary condition) is solvable.
Furthermore it is known that a infinite set of eigen-values $\{ \lambda_i
\}_{i=0}^{\infty}$ of $-\nabla^2$ satisfy
\begin{itemize}
 \item $0 < \lambda_0  < \lambda_1  \le \lambda_2 \le \cdots$; 
 \item $\lambda_0$ is a simple eigen-value.
\end{itemize}
Let the eigen-function corresponding to the eigen-value $\lambda_0$ be $u_0$ (i.e., $-\nabla^2 u_0 = \lambda_0 u_0$).
First, according to the simple property of the eigen-value $\lambda_0$, it is clear that
\[ {\rm Ker}(f_{u}(\lambda_0,0)) = \{ t u_0;~ t \in R^1 \},  \]
so that the dimension of ${\rm Ker}(f_{u}(\lambda_0,0))$ is equal to 1.
Second, if there exists a solution $v \in X$ for $\nabla^2 v - \lambda_0 v = h$
with $h \in Y$, then
\[ R(f_{u}(\lambda_0,0)) = \left\{ h \in Y; \int_{\Omega} h(x) u(x) dx = 0 \right\},  \]
so that $R(f_{u}(\lambda_0,0))$ is a closed subset of $Y$ with its co-dimension 1 (cf. the Riesz-Schauder theory~\cite{80yosida}).
Third, it is valid that 
\begin{equation} 
f_{u \lambda}(\lambda_0, 0) [u]  =  - \lambda_0 u  \notin R(f_{u}(\lambda_0, 0)).
\end{equation}
Consequently, according to the bifurcation theory~\cite{71crandall, 73crandall}, $\lambda = \lambda_0$ has
been clarified to be a bifurcation point (corresponding to ($\lambda_0,0$) in Fig.~1).
Note that only sufficient conditions for
the existence of the bifurcation point is presented in the bifurcation theory.

\begin{figure} \label{fig1}
\begin{center}
\includegraphics[width=9pc]{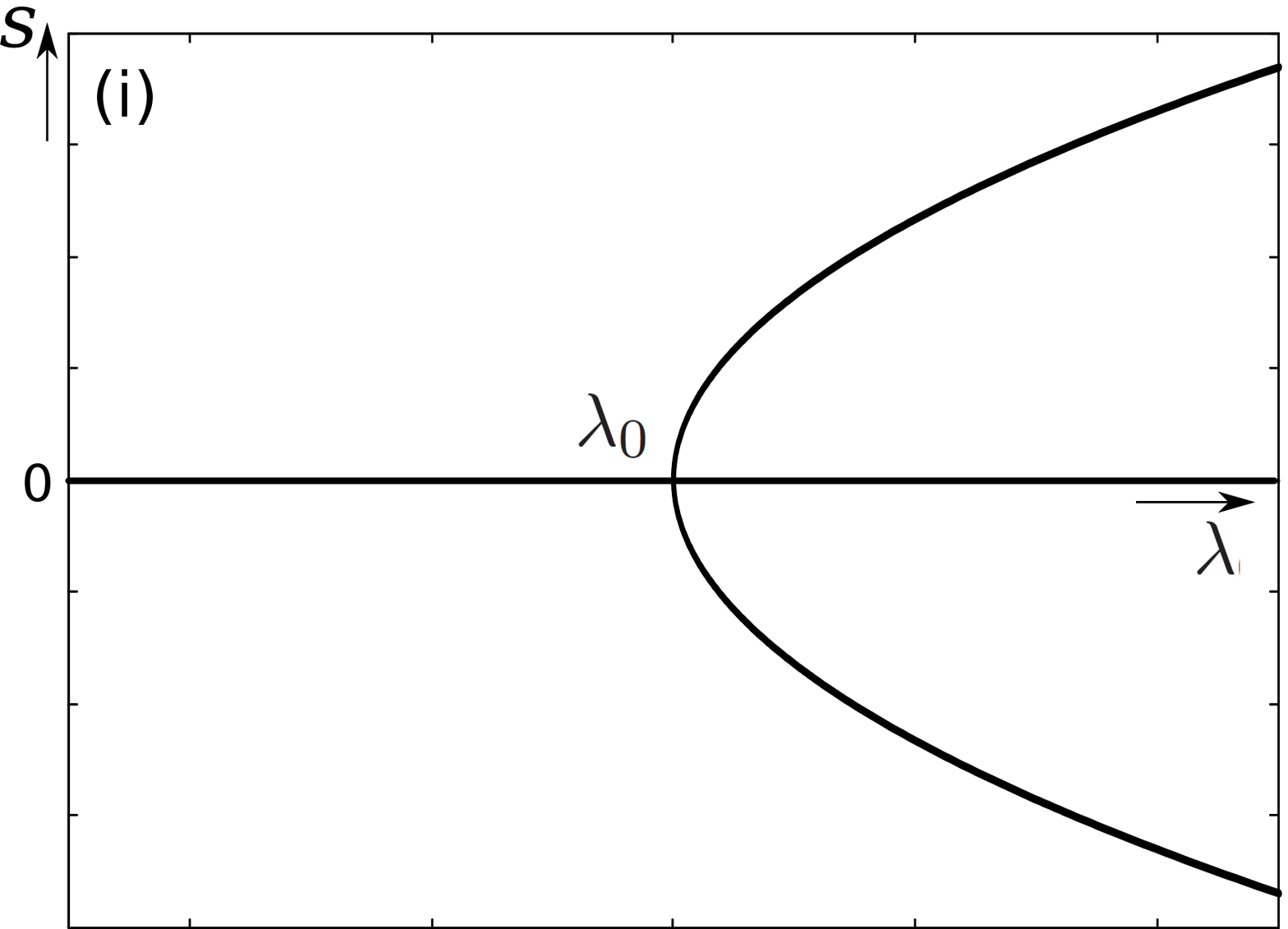}\hspace{2pc}
\includegraphics[width=9pc]{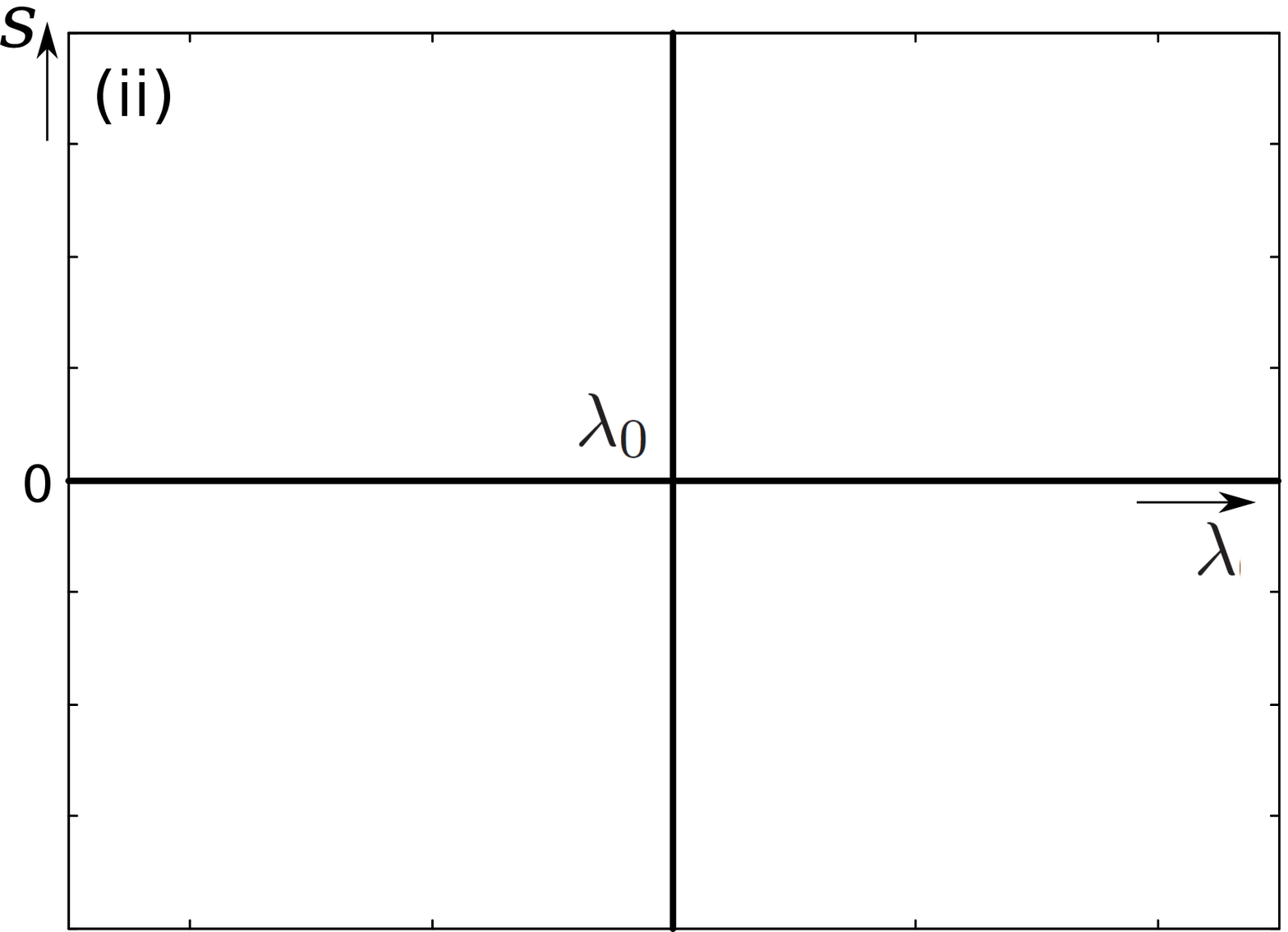}\hspace{2pc} 
\includegraphics[width=9pc]{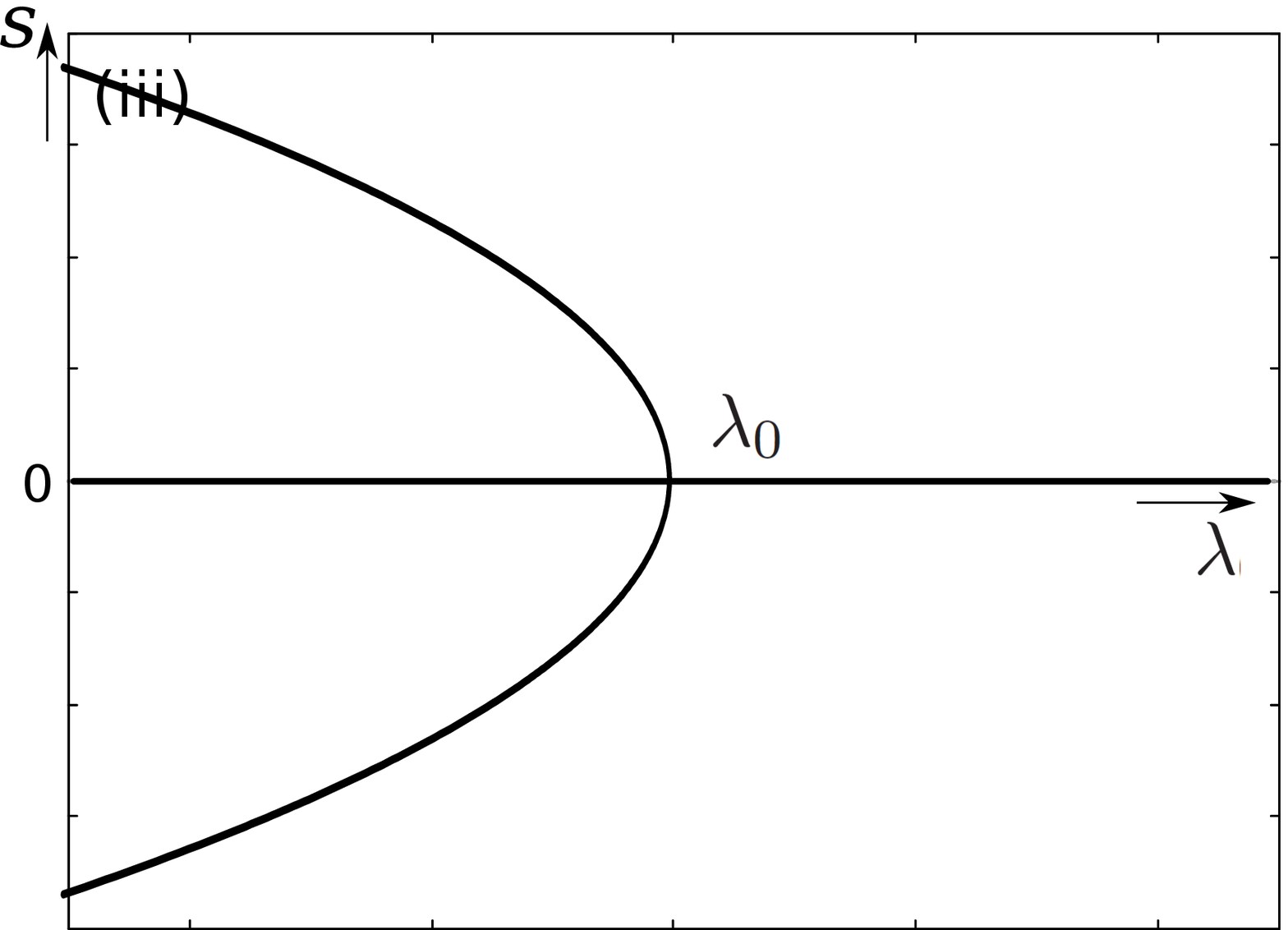}\hspace{2pc} \\
\includegraphics[width=9pc]{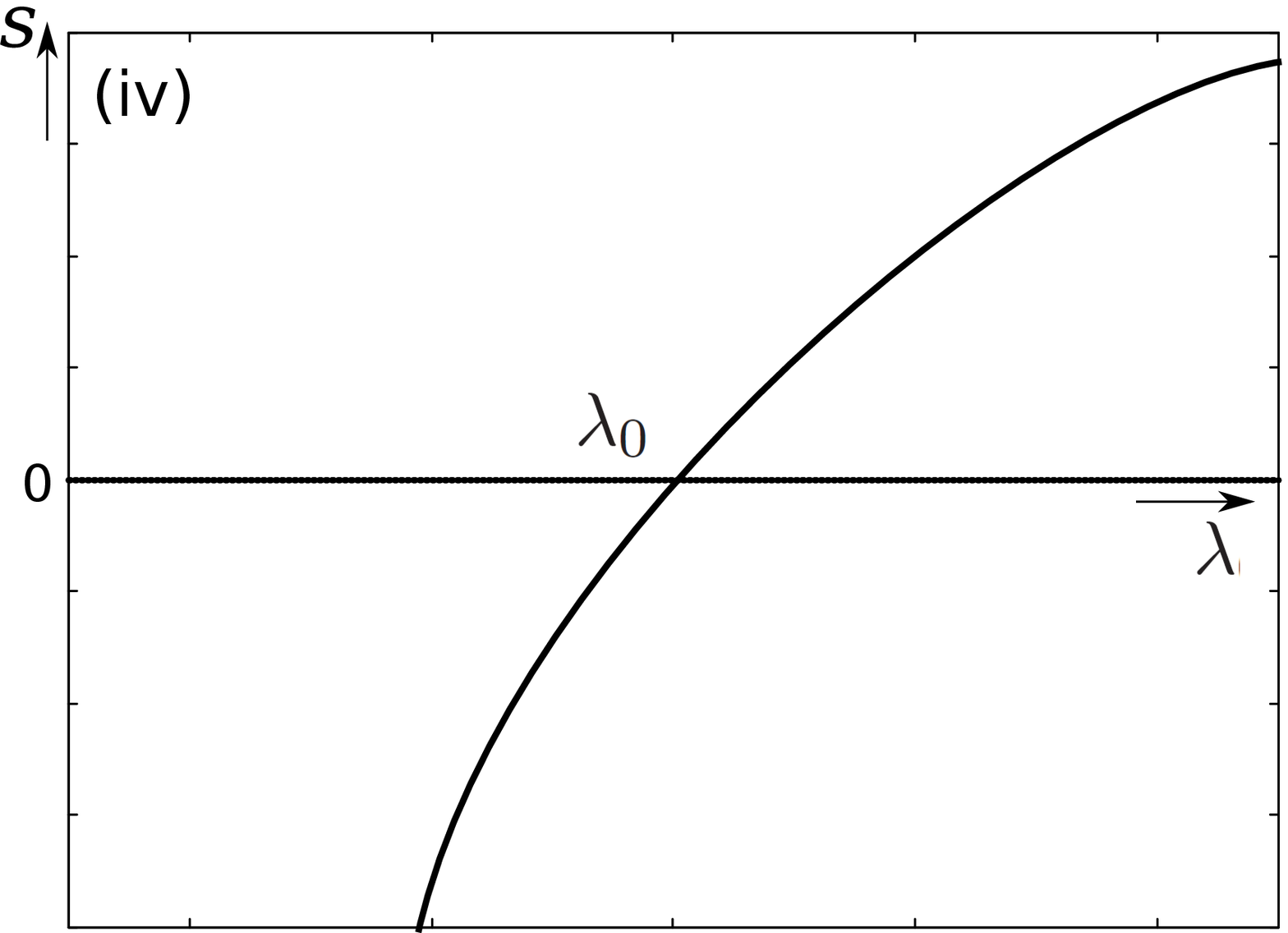}\hspace{2pc}
\includegraphics[width=9pc]{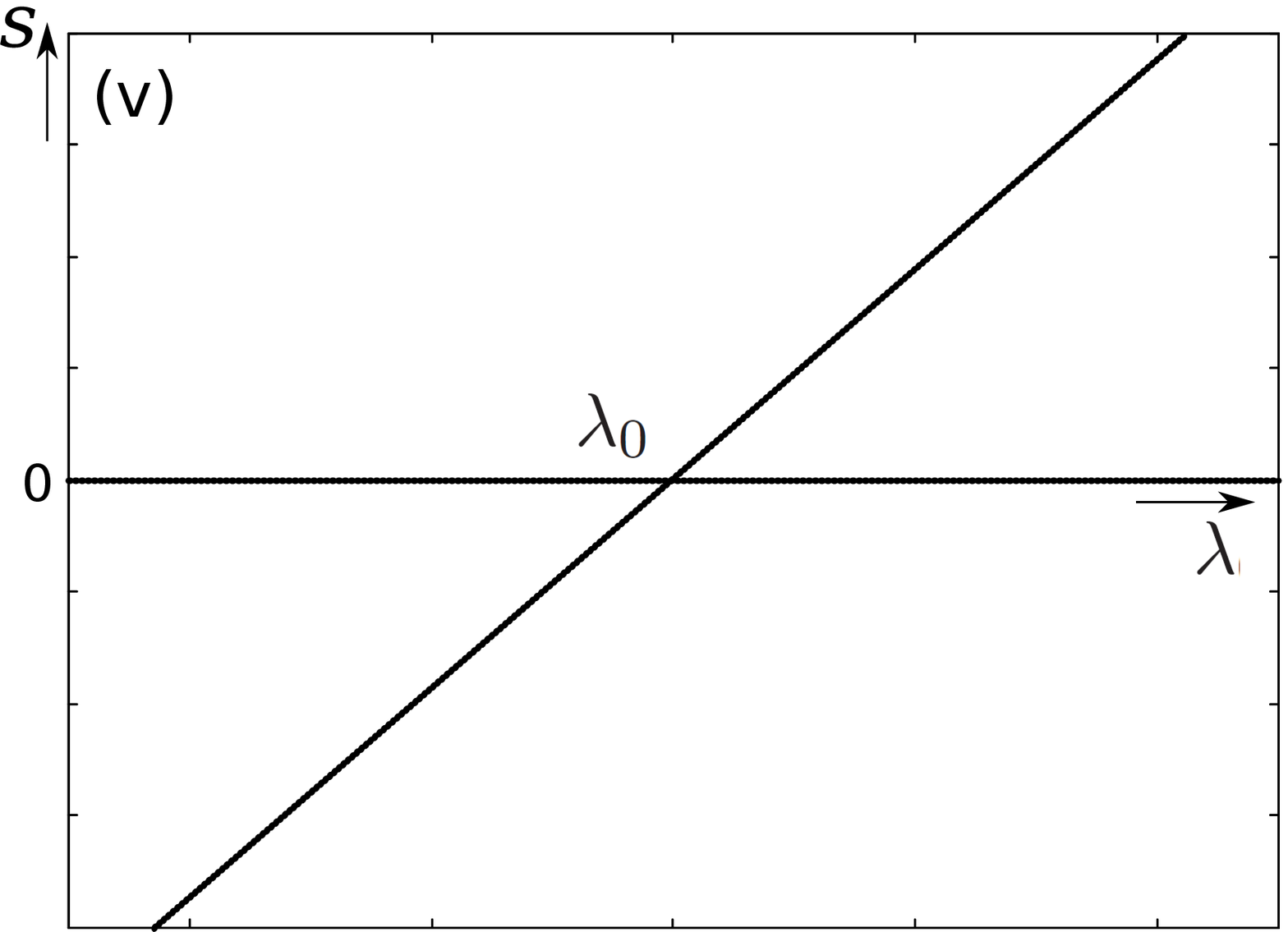}\hspace{2pc}
\includegraphics[width=9pc]{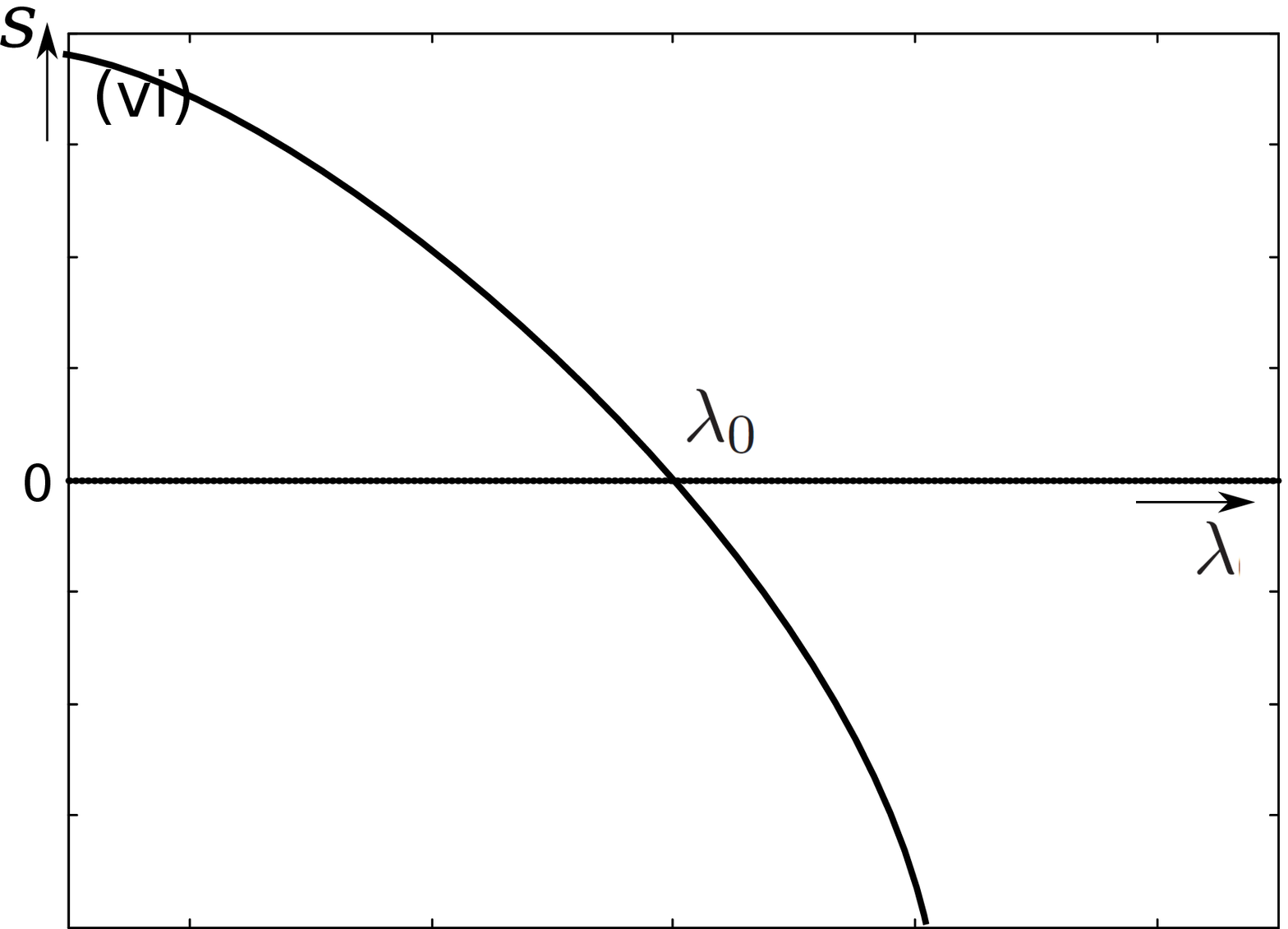}\hspace{2pc} \\
\includegraphics[width=9pc]{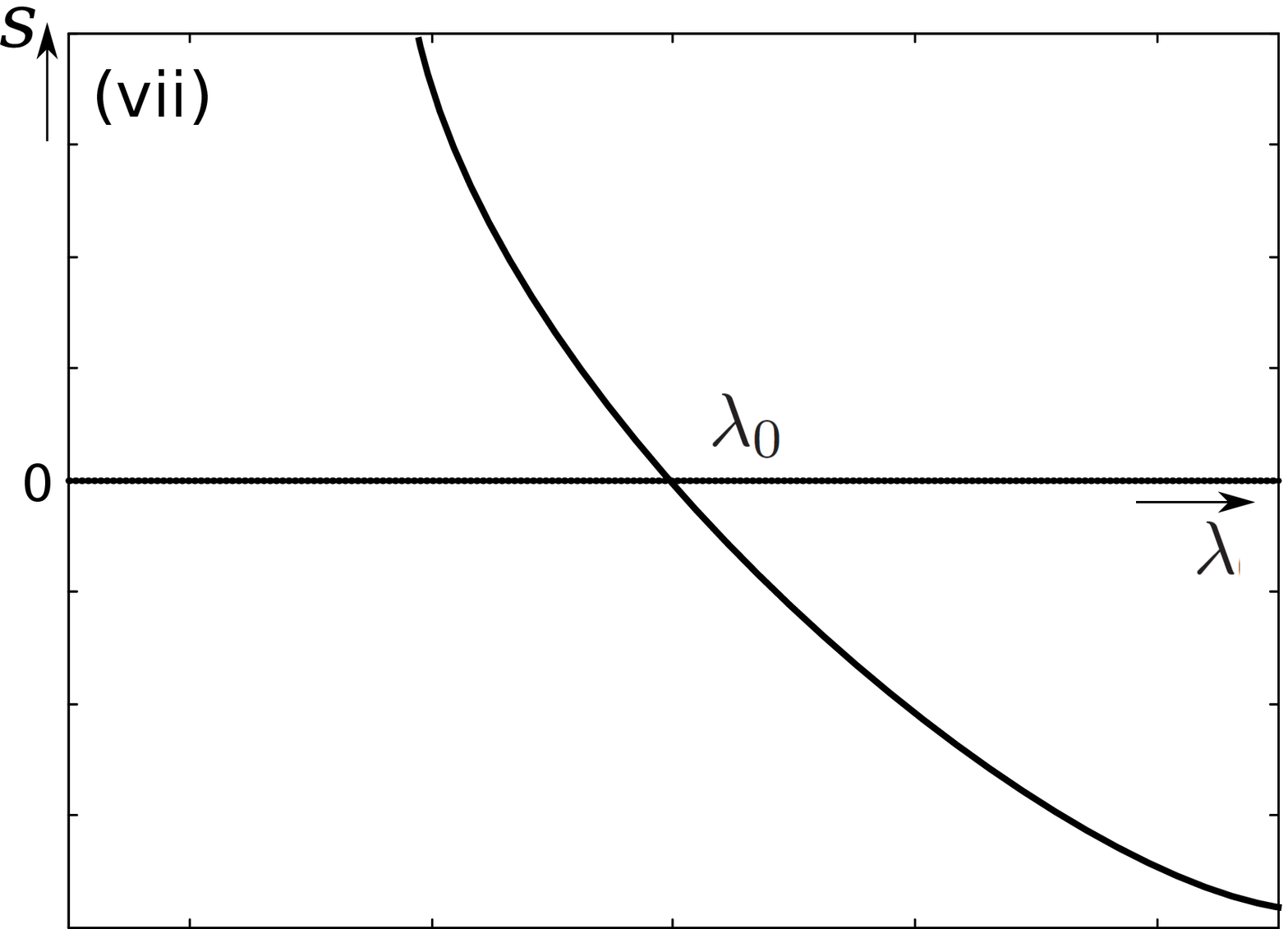}\hspace{2pc}
\includegraphics[width=9pc]{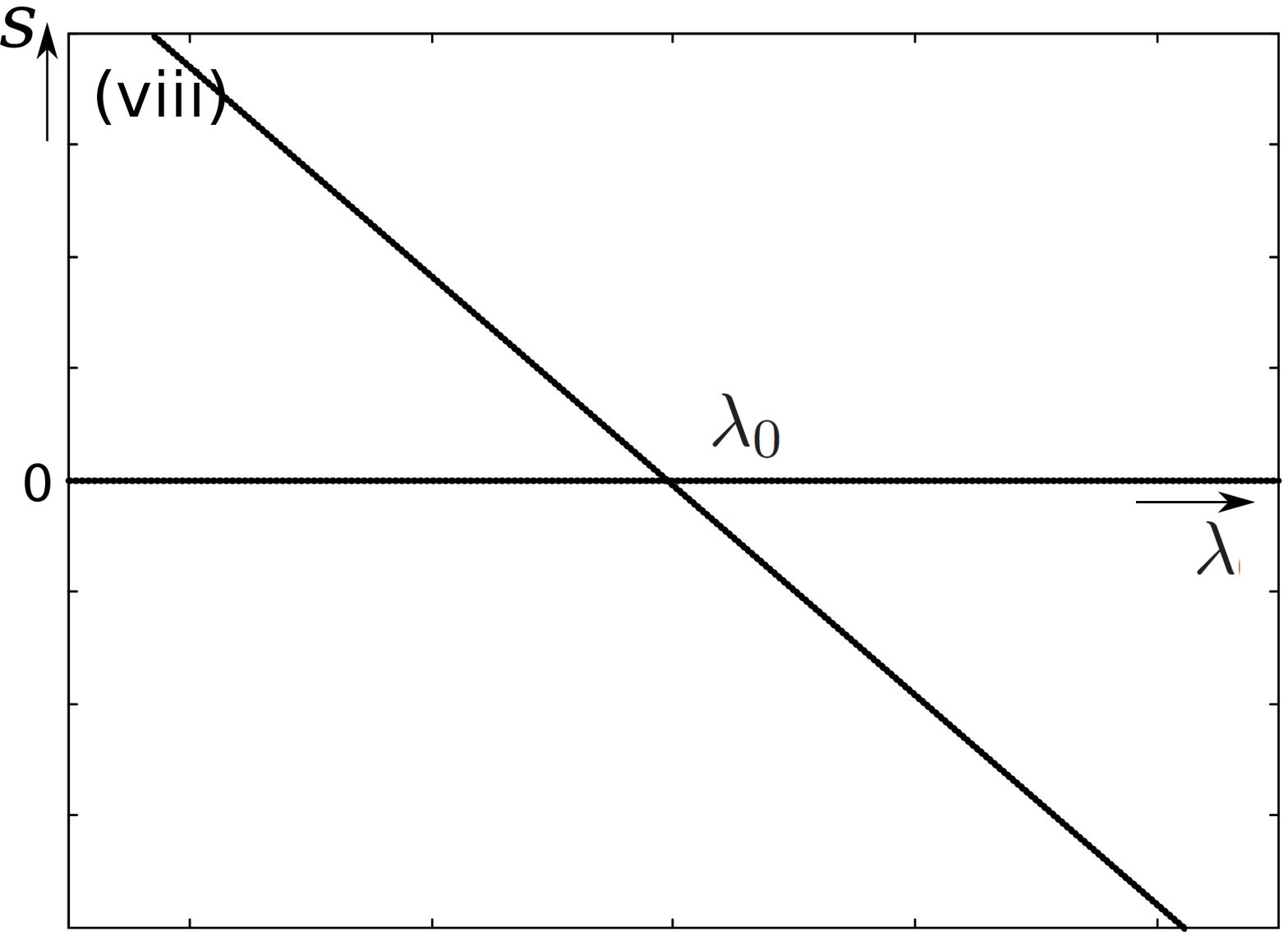}\hspace{2pc} 
\includegraphics[width=9pc]{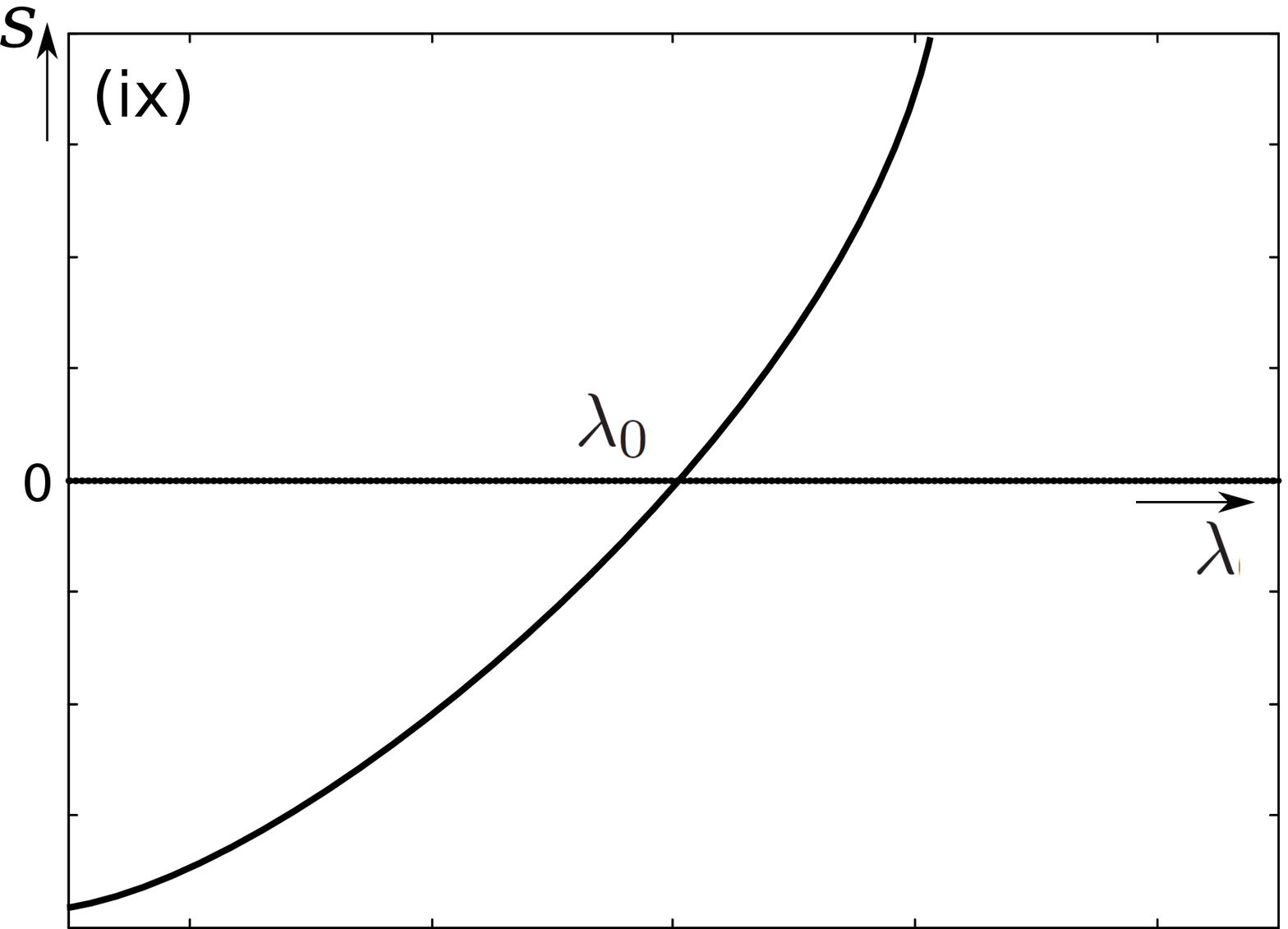}\hspace{2pc} \\
\end{center}
\caption{\label{label}
9 types of co-existence based on Eqs.~(\ref{res1}) and
(\ref{res2}): cases (i), (ii), and
(iii) appear if $\mu_s(0) =0$, cases (iv), (v),
and (vi)) appear if $\mu_s(0) > 0$, cases (vii),
(viii), and (ix) appear if $\mu_s(0) < 0$;
cases (i), (iv), and (vii) appear if $\mu_{ss}(0) > 0$, cases (ii), (v), and (viii) appear if $\mu_{ss} = 0$,
cases (iii), (vi), and (ix) appear if $\mu_{ss} < 0$. }
\end{figure}

\subsection{Nonlinear analysis}
Co-existence of different states (i.e., existence of non-trivial solution $u \ne 0$) is shown.
We set a closed interval $[-\epsilon_0, \epsilon_0]$ and a $C^1$-function
$\lambda(s)$ satisfying $\lambda(0)= \lambda_0$, where $s$ parametrizes the functional space $X$.
Under the three conditions confirmed in Sec.~\ref{linear}, let the corresponding solution $u$ be represented by
\[ u(\lambda,s,x) = s u_0(\lambda,x)+ s z(\lambda,s,x),   \]
where $s$ is defined on the interval, and $z(\lambda,s,x)$ is a sufficiently smooth function of $s$
defined on $R^1 \times R^1 \times X$.
The function $z(\lambda,s,x)$ satisfies $z(\lambda,0,x)=0$ and
\[ \int_{\Omega} z(x) u_0(x) dx = 0. \]
The function $u(\lambda,s,x)$ satisfies the condition $u(\lambda,0,x)= 0$, which means the existence of the trivial
solution.
It is useful to define a linear operator
\[ 
A := -\nabla^2 - \lambda_0,
 \]
with its domain $X$, and then it is readily seen that $A$ is a self-adjoint operator in $Y$.
The original equation is written by $A u = \mu(s) u + (V(u) -V_L) u$
with $\mu(s) = \lambda(s)- \lambda_0$, and the linearized problem is written by $A u_0 =0$.
By differentiating the original equation with respect to $s$, step by step
\[ \begin{array}{ll}
(A u)_s = \mu_s u +  \mu u_s  + \partial_s (V(u)u) -V_L u_s \vspace{1.5mm} \\
 \quad = \mu_s u +  \mu u_s +  (\partial_s V(u)) u +
 V(u)u_s -V_L u_s \vspace{2.5mm} \\
(A u)_{ss} = \mu_{ss} u + 2 \mu_s u_s + \mu u_{ss} +
  \partial_s^2 (V(u)u) -V_L u_{ss}  \vspace{1.5mm} \\
\quad = \mu_{ss} u + 2 \mu_s u_s + \mu u_{ss} 
 +  (\partial_s^2 V(u)) u + 2 (\partial_s V(u)) u_s  +
 V(u)u_{ss}  -V_L u_{ss}   \vspace{2.5mm} \\
(A u)_{sss} = \mu_{sss} u  + 3 \mu_{ss} u_s + 3
  \mu_s u_{ss} +  \mu u_{sss} +  \partial_{s}^3
  (V(u)u)  -V_L u_{sss} \vspace{1.5mm} \\
\quad = \mu_{sss} u  + 3 \mu_{ss} u_s + 3
  \mu_s u_{ss} +  \mu u_{sss} 
+  (\partial_s^3 V(u)) u 
+ 3 (\partial_s^2 V(u)) u_s + 3 (\partial_s V(u)) u_{ss}  \vspace{1.5mm} \\
\quad +  V(u)u_{sss}   -V_L u_{sss}
\end{array} \]
where the functions are represented by  
$u_s =u_0 + s z_s + z$, $u_{ss} = s z_{ss} + 2 z_s$, and $u_{sss}
=  s z_{sss} + 3 z_{ss}$ respectively.
The derivatives of the inhomogeneous terms become
\[ \begin{array}{ll} 
\partial_s V(u) = (\partial_{u} V(u)) ~  u_s  \vspace{1.5mm} \\
\partial_s^2 V(u)  = (\partial_{u}^2 V(u)) ~  u_s^2  +
(\partial_{u} V(u)) ~  u_{ss}  \vspace{1.5mm} \\
\partial_s^3 V(u)  = (\partial_{u}^3 V(u)) ~  u_s^3  + 3
(\partial_{u}^2 V(u)) ~  u_s u_{ss} 
 + (\partial_{u} V(u)) ~  u_{sss}.
 \end{array} \]
By taking $s=0$, the bi-linear forms become
\[ \begin{array}{ll}
 (A u)_{s}|_{s=0} = \mu_s(0) u|_{s=0} +  \mu(0) u_s|_{s=0} +
  \partial_{u}( V(u)  u) u_s|_{s=0} - V_L  u_s |_{s=0}  \vspace{1.5mm} \\
\quad  =   V_L  u_s|_{s=0} - V_L  u_s|_{s=0} = 0, \vspace{1.5mm} \\
 ((A u)_{s}|_{s=0}, u_0)  = 0,  \vspace{5mm} \\
 (A u)_{ss}|_{s=0} =  \mu_{ss}(0) u|_{s=0} + 2 \mu_s(0) (u_0  +
z|_{s=0}) + 2 \mu(0) z_{s}|_{s=0} +  \partial_s^2 (V(u) u)|_{s=0} - 2 V_L  z_s|_{s=0} \vspace{1.5mm} \\
\quad = 2 \mu_s(0) u_0   +  \partial_s^2 (V(u) u)|_{s=0} - 2 V_L  z_s|_{s=0}, \vspace{1.5mm} \\
 ((A u)_{ss}|_{s=0}, u_0) = 2 \mu_s(0)  ( u_0, u_0)  +
 (\partial_s^2 (V(u) u)|_{s=0}, u_0) - (2 V_L ~  z_s|_{s=0},u_0),  \vspace{5mm} \\
 (A u)_{sss}|_{s=0} =  \mu_{sss}(0) u|_{s=0}  + 3 \mu_{ss}(0)
  (u_0 + z|_{s=0} + 6
  \mu_s(0) z_{s}|_{s=0} + 3 \mu(0) z_{ss}|_{s=0} \vspace{1.5mm} \\
\quad +  \partial_s^3
  (V(u)u)|_{s=0} - 3 V_L ~  z_{ss}|_{s=0} \vspace{1.5mm} \\
\quad  = 3 \mu_{ss}(0) u_0  +  6
  \mu_s(0) z_{s}|_{s=0}  
+  \partial_s^3
  (V(u)u)|_{s=0}  - 3 V_L ~  z_{ss}|_{s=0},  \vspace{1.5mm} \\
 ((A u)_{sss}|_{s=0}, u_0) = 3 \mu_{ss}(0)  ( u_0, u_0)
+  (6 \mu_s(0) z_{s}|_{s=0},u_0)  
+ (\partial_s^3(
  V(u)u)|_{s=0},u_0)  - 3 (V_L ~  z_{ss}|_{s=0},u_0),
\end{array} \]
where $u|_{s=0}= 0$, $z|_{s=0}=0$, and $\mu(0)=0$ are utilized, as well
as Eq.~(\ref{inhomcond}).
$(A u_{ss}|_{s=0}, u_0) = (u_{ss}|_{s=0},
  A u_0)  = 0$ due to $A
  u_0 =0$.
Consequently
\begin{equation} \label{res1} \begin{array}{ll}
 2 \mu_s(0)   = - (\partial_s^2 (V(u) u)|_{s=0}, u_0) +  2( V_L   z_s|_{s=0},u_0),
\end{array} \end{equation}
and the sign of $\lambda_s(0) = \mu_s(0)$ is determined by $-(\partial_s^2
(V(u) u)|_{s=0}, u_0)+  2( V_L   z_s|_{s=0},u_0)$.
In the same manner $(A u_{sss}|_{s=0}, u_0) = (u_{sss}|_{s=0},
  A u_0)  = 0$.
It leads to
\begin{equation} \label{res2} \begin{array}{ll}
  3 \mu_{ss}(0) =  -(\partial_s^3 ( V(u) u)|_{s=0},u_0) -  (6
  \mu_s(0) z_{s}|_{s=0},u_0)  + 3 (V_L  z_{ss}|_{s=0},u_0), 
\end{array} \end{equation}
and the sign of $\lambda_{ss}(0) = \mu_{ss}(0)$ is determined by $ -(\partial_s^3
(V(u)u )|_{s=0},u_0) -  (6 \mu_s(0) z_{s}|_{s=0},u_0)+ 3 (V_L  z_{ss}|_{s=0},u_0)$.
In particular, if $\lambda_s(0) = \mu_s(0) = 0$ is true, the sign of $\lambda_{ss}(0)$ is determined by $ -(\partial_s^3
(V(u)u )|_{s=0},u_0)+ 3 (V_L  z_{ss}|_{s=0},u_0)$.
According to Eqs.~(\ref{res1}) and (\ref{res2}), the co-existence of states is classified into 9 types (Fig.~1). 
In Figure~1, around the neighbour of the bifurcation point $(\lambda_0,0)$,
two solutions co-exist in types (iv) to (ix), while the transition from
single-existence to co-existence is described in types (i) and (iii).

\begin{table}[t] \label{table}
\caption{Systematic analysis for $\psi^k$-interaction theory.  
Possible classification of co-existence is shown in the column ``Type'', where
$\sigma = 4\eta   ( u_0  z_s|_{s=0} ,u_0) -  2
  \eta  (u_0^2, u_0) ( z_{s}|_{s=0},u_0) $.}
\begin{center}
  \begin{tabular}{|l||c|c||c|c||c|c||c||} \hline
    $k$ &  $\partial_s V(u)$ & $\partial_s^2 V(u)$  & $\partial_s^2 (V(u) u)|_{s=0}$ & $ \partial_s^3 (V(u)
    u)|_{s=0}$ & $\mu_s(0)$ & $\mu_{ss}(0)$ & Type  \\ \hline \hline
   = 3 & $-\eta u_s$  & $-\eta u_{ss}$ &
$ - 2 \eta  u_0^2 $ & 
$- 12 \eta u_0  z_{s}|_{s=0}$ &  $ \eta  (u_0^2, u_0)$ &     $\sigma$     & all  \\ \hline
   = 4 &  0 & $-2 \eta u_{s}^2$  & $0$ & $-6 \eta  u_{0}^3 $ &  0
   & $2 \eta  (u_{0}^3, u_0) $  & (i),(ii),(iii) \\ \hline
   $\ge$ 5 & 0  & 0 & 0  & 0 &  0 & 0   & (ii) \\ \hline
  \end{tabular}
\end{center}
\end{table}

\section{Application to $ \psi^k$-interaction theory}
If the Lagrangian includes the $k$th-order nonlinearity in its interaction part (for example, see
textbooks of particle physics),
the inhomogeneous term of the master equation becomes 
\[ V(u) u = - \eta u^{k-1}, \]
for integers $k \ge 1$, where $\eta$ is assumed to be a real number.
Here $V_L = 0$ and $V(u)|_{s=0} = V(u)|_{u=0} = 0$ are true.
The first derivative is
\[  \begin{array}{ll} 
\partial_{u} V(u)|_{u=0} = - (k-2) \eta u^{k-3}|_{u=0}
\end{array} \]
for $k \ge3$, so that it is equal to $- \eta $ for $k=3$, and zero for $k \ge 4$.
The second derivative is
\[  \begin{array}{ll} 
\partial_{u}^2 V(u)|_{u=0} = - (k-2)(k-3) \eta u^{k-4}|_{u=0}
\end{array} \]
for $k \ge4$, so that it is equal to zero for $k=3$, $-2 \eta$ for $k=4$, and zero for $k \ge 5$. 
The third derivative is
\[  \begin{array}{ll} 
\partial_{u}^3 V(u)|_{u=0} = - (k-2)(k-3)(k-4) \eta u^{k-5}|_{u=0}
\end{array} \]
for $k \ge5$, so that it is equal to zero for $k \le 4$, $-6 \eta$ for $k=5$,
and zero for $k \ge 6$.
Results are summarized in Table~1. 
In case of $k=4$ ($\psi^4$-interaction theory), the non-trivial solution
corresponds to type (i) of Fig.~1 if $\eta > 0$, to type (ii) if $\eta = 0$,
and to type (iii) if $\eta < 0$.
In particular when $\eta >0$, the co-existence emerges only if $m > \lambda_0$
(cf. spontaneous symmetry breaking). 

If there is no interaction (free particle condition: $\eta=0$), $\mu_s(0)= \mu_{ss}(0)= 0$ is
true, and the co-existence is classified into type (ii).
If the interaction is linear ($V(u)= V_L \ne 0$; $\psi^2$-interaction theory),  the derivatives are
\[  \begin{array}{ll} 
\partial_{u} V(u)|_{u=0} = \partial_{u}^2 V(u)|_{u=0} = \partial_{u}^3 V(u)|_{u=0} = 0,
\end{array} \]
so that  $\partial_s V(u) = \partial_s^2 V(u) = \partial_s^3 V(u) =
0$.
It leads to $\partial_s^2 (V(u) u)|_{s=0} = V_L u_{ss}|_{s=0}$ and
$\partial_s^3 (V(u)    u)|_{s=0} = V_L u_{sss}|_{s=0}$ so that
$\mu_s(0)   = - (  V_L z_s|_{s=0} , u_0) +  ( V_L z_s |_{s=0},u_0) = 0$
and
$ \mu_{ss}(0) =  -( V_L  z_{ss}|_{s=0},u_0)  +  (V_L z_{ss}|_{s=0},u_0) = 0$
follows.
The co-existence is classified into type (ii).
As a result the nonlinearity can be identified by the classification other than type (ii).

\ack{
This work was supported by HPCI Strategic Programs for Innovative Research Field 5 ``The origin of matter and the universe".
The author is grateful to Prof. Emeritus Dr. Hiroki Tanabe for reading the manuscript.
}

\end{document}